# Energy- and flux-budget turbulence closure model for stably stratified flows. Part II: the role of internal gravity waves

S. S. Zilitinkevich[1,2,3], T. Elperin[4], N. Kleeorin[4], V. L'vov[5] and I. Rogachevskii[4]

[1]Finnish Meteorological Institute, Helsinki, Finland

[2]Division of Atmospheric Sciences, University of Helsinki, Finland

[3]Nansen Environmental and Remote Sensing Centre / Bjerknes Centre for Climate Research, Bergen, Norway

[4]Pearlstone Centre for Aeronautical Engineering Studies, Department of Mechanical Engineering, Ben-Gurion University of the Negev, Beer-Sheva, Israel

[5]Department of Chemical Physics, Weizmann Institute of Science, Israel

## Abstract

We advance our prior energy- and flux-budget (EFB) turbulence closure model for the stably stratified atmospheric flows and extend it accounting for additional vertical flux of momentum and additional productions of turbulent kinetic energy (TKE), turbulent potential energy (TPE) and turbulent flux of potential temperature due to large-scale internal gravity waves (IGW). For the stationary, homogeneous regime, the first version of the EFB model disregarding large-scale IGW yielded universal dependencies of the flux Richardson number, turbulent Prandtl number, energy ratios, and normalised vertical fluxes of momentum and heat on the gradient Richardson number, Ri. Due to the large-scale IGW, these dependencies lose their universality. The maximal value of the flux Richardson number (universal constant $\approx$ 0.2-0.25 in the no-IGW regime) becomes strongly variable. In the vertically homogeneous stratification, it increases with increasing wave energy and can even exceed 1. In the heterogeneous stratification, when IGW propagate towards stronger stratification, the maximal flux Richardson number decreases with increasing wave energy, reaches zero and then becomes negative. In other words, the vertical flux of potential temperature becomes counter-gradient. IGW also reduce anisotropy of turbulence: in contrast to the mean wind shear, which generates only horizontal TKE, IGW generate both horizontal and vertical TKE. IGW also increase the share of TPE in the turbulent total energy (TTE = TKE + TPE). A well-known effect of IGW is their direct contribution to the vertical transport of momentum. Depending on the direction (downward or upward), IGW either strengthen or weaken the total vertical flux of momentum. Predictions from the proposed model are consistent with available data from atmospheric and laboratory experiments, direct numerical simulations and large-eddy simulations.

**Key words:**  Internal gravity waves (IGW)   Stable stratification   Turbulence closure   Turbulent energies   Vertical turbulent fluxes   Wave-induced transports

# List of Symbols

$A_z = E_z/E_K$ is the ratio of the vertical turbulent kinetic energy, $E_z$, to TKE, $E_K$

$E = E_K + E_P$ is the total turbulent energy (TTE)

$E_K = \frac{1}{2}\langle u_i u_i \rangle$ is the turbulent kinetic energy (TKE)

$E_i$ are the vertical $(i = z)$ and horizontal $(i = x, y)$ components of TKE

$E_\theta = \frac{1}{2}\langle \theta^2 \rangle$ is the "energy" of potential temperature fluctuations

$E_P$ is the turbulent potential energy (TPE) given by Eq. (25)

$E_W$ is the IGW kinetic energy given by Eq. (16)

$\hat{E}_z$ is the dimensionless vertical TKE given by Eq. (70)

**e** is the vertical unit vector

$e_W(\mathbf{k})$ is the energy spectrum of the ensemble of internal gravity waves (IGW) given by Eq. (17)

$F_i = \langle u_i \theta \rangle$ is the potential-temperature flux

$F_z$ is the vertical component of the potential-temperature flux

$F_i^W$ is the instantaneous potential temperature flux caused by the IGW-turbulence interaction given by Eq. (42)

$F_i^{WW}$ is $F_i^W$ averaged over the period of IGW, given by Eq. (21)

$F_\theta^W$ is a scale of the IGW contribution turbulent flux of potential temperature given by Eq. (57)

$\hat{E}_z$ is the dimensionless vertical potential-temperature flux given by Eq. (70)

$f = 2\Omega \sin \varphi$ is the Coriolis parameter

$G$ is a "wave-energy parameter" proportional to the normalized IGW kinetic energy, $E_W$, given by Eq. (50)

**g** is the acceleration due to gravity

$H$ is the external height scale

$K_M$ is the eddy viscosity given by Eq. (79)

$K_H$ is the eddy conductivity given by Eq. (80)

**k** is the wave vector

$k_\alpha = (k_x, k_y)$ is the horizontal wave vector with $k_h = \pm\sqrt{k_x^2 + k_y^2}$

$k = \sqrt{k_z^2 + k_h^2}$ is the total wave number

$L$ is the Monin-Obukhov length scale given by Eq. (5)

$L_W$ is the minimal wave length of the large-scale IGW

$l_z$ is the vertical turbulent length scale

$N$ is the mean-flow Brunt-Väisälä frequency

$P$ is the pressure

$P_0$ is the reference value of $P$

$P^W$ is the pressure variation caused by IGW

$p$ is the pressure fluctuation caused by turbulence

$\Pr = \nu/\kappa$ is the Prandtl number

$\Pr_T$ is the turbulent Prandtl number given by Eq. (2)

$Q$ is the dimensionless lapse rate given by Eq. (44)

$Q_{ij}$ are correlations between the pressure and the velocity-shear fluctuations, given by Eqs. (34), (63)

$\mathbf{r}$ is the radius-vector of the centre of the wave packet

$\mathrm{Ri}$ is the gradient Richardson number, Eq. (1)

$\mathrm{Ri}_f$ is the flux Richardson number, Eq. (4)

$\mathrm{Ri}_f^\infty$ is the limiting value of $\mathrm{Ri}_f$ in the flow without IGW (universal constant in the homogeneous sheared flow)

$\mathrm{Ri}_f^{\max}$ is the limiting value of $\mathrm{Ri}_f$ in the presence of IGW (depends on the $G$ and $Q$)

$S = |\partial \mathbf{U}/\partial z|$ is the vertical shear of the horizontal mean wind

$T$ is the absolute temperature

$T_0$ is the reference value of the absolute temperature

$t_T = l_z E_z^{-1/2}$ is the turbulent dissipation time scale

$t_\tau$ is the effective dissipation time scale

$\mathbf{V}^W = (V_1^W, V_2^W, V_3^W)$ is the IGW velocity given by Eqs. (9)-(10)

$V_0^W(\mathbf{k})$ is the IGW amplitude

$\mathbf{U} = (U_1, U_2, U_3)$ is the mean wind velocity

$\mathbf{u}$ is the turbulent velocity

$Z_0$ is the height of the IGW source

$\beta = g/T_0$ is the buoyancy parameter

$\gamma = c_p/c_v$ is the specific heats ratio

$\varepsilon_K$, $\varepsilon_\theta$, $\varepsilon_i^{(F)}$ and $\varepsilon_{ij}^{(\tau)}$ are the dissipation rates for $E_K$, $E_\theta$, $F_i^{(F)}$ and $\tau_{ij}$

$\varepsilon_{\alpha 3(\mathrm{eff})}$ ($\alpha = 1,2$) are the effective dissipation rates of the vertical turbulent fluxes of momentum

$\kappa$ is the temperature conductivity

$\mu$ is the exponent of the energy spectrum of the ensemble of IGW

$\nu$ is the kinematic viscosity

$\Phi_K$, $\Phi_\theta$ and $\Phi_F$ are the third-order moments representing turbulent fluxes of $E_K$, $E_\theta$ and $F_i$

$\varphi$ is the latitude

$\Pi^W$ is the IGW production of TKE given by Eq. (49)

$\Pi_i^W$ is the IGW production of the vertical ($i = z$) and horizontal ($i = x, y$) components of TKE given by Eqs. (52)-(53)

$\Pi_\theta^W$ is the IGW production of $E_\theta$ given by Eq. (56)

$\Pi_E^W = \Pi^W + \Pi_\theta^W \beta^2 N^{-2}$ is the IGW production of TTE

$\Pi_F^W$ is the IGW production of the flux of potential temperature given by Eq. (58)

$\Pi^W_{\tau\alpha}$ is the IGW production of the non-diagonal components of the Reynolds stresses, $\tau_{\alpha 3}$, given by Eq. (61)

$\tau_{ij}$ are the Reynolds stresses characterising the turbulent flux of momentum

$\tau_{\alpha 3}$ ($\alpha = 1,2$) are the components of the Reynolds stresses characterising the vertical turbulent flux of momentum

$\tau$ is the modulus of ($\tau_{13}, \tau_{23}$)

$\tau^W_{ij}$ are the instantaneous Reynolds stresses caused by IGW, given by Eq. (41)

$\tau^{WW}_{ij}$ are the Reynolds stresses caused by IGW averaged over the period of IGW given by Eqs. (21), (43)

$\rho_0$ is the density

$\Theta$ is the potential temperature

$\Theta^W$ is the IGW potential temperature given by Eq. (11)

$\theta$ is the turbulent fluctuation of potential temperature

$\Omega_i$ is the Earth's rotation vector parallel to the polar axis

$\omega$ is the frequency of IGW

# 1. Introduction

Internal gravity waves (IGW) in relation to atmospheric and oceanic turbulence have been a subject of intense research [e.g., monographs: Beer (1974), Gossard and Hooke (1975), Baines (1995), Nappo (2002); and review papers: Garrett and Munk (1979), Fritts and Alexander (2003), Thorpe (2004), Staquet and Sommeria (2002)]. In the atmosphere, IGW are present at scales ranging from meters to kilometers, and are measured by direct probing or remote sensing using radars and lidars (Chimonas, 1999; Fritts and Alexander 2003).

The sources of IGW are: strong wind shears, flows over topography, convective and other local-scale motions underlying the stably stratified layer (Wurtele et al., 1996; Fritts and Alexander, 2003), geostrophic adjustment of unbalanced flows in the vicinity of jet streams and frontal systems, and wave-wave interactions (Staquet and Sommeria, 2002; Fritts and Alexander, 2003). The IGW propagation is complicated by variable wind and density profiles causing refraction, reflection, focusing, and ducting.

IGW contribute to the energy and momentum transport, the turbulence production and eventually enhance mixing. The dual nature of fluctuations representing both turbulence and waves in stratified flows has been recognised, e.g., by Jacobitz et al. (2005). The role of waves and the need for their inclusion in turbulence closure models has been discussed by Jin et al. (2003) and Baumert and Peters (2004, 2009). Baumert and Peters (2004) included an additional negative term in the TKE budget equation: the rate of transfer of TKE into potential energy of wave-like motions (highly irregular short internal waves coexisting with turbulent eddies); and postulated that with increasing stability these motions dominate random velocity and buoyancy fluctuations and suppress vertical mixing [see also Umlauf and Burchard (2005)]. Parameterization of mixing in the deep ocean due to short IGW was considered by Polzin (2004a, 2004b). Finnigan and Einaudi (1981), Einaudi et al. (1984), Finnigan et al. (1984), Finnigan (1988, 1999), Einaudi and Finnigan (1993) analyzed the budgets of the wave kinetic energy and the wave temperature variance. They found significant buoyant production of the wave energy despite the strong static stability and energy transfer from waves to turbulence.

In the present paper we focus on the wave-induced vertical flux of momentum and the generation of turbulent kinetic energy (TKE), turbulent potential energy (TPE) and turbulent flux of potential temperature due to large-scale IGW in the context of an energetically consistent "energy- and flux-budget" (EFB) turbulence closure model for the stably stratified flows (Zilitinkevich et al., 2007, 2008). The model is designed for typical stably stratified atmospheric flows, characterised by the vertical shear $S = |\partial \mathbf{U}/\partial z|$ of the horizontal mean wind $\mathbf{U} = (U_1, U_2, 0)$, and is based on the budget equations for the key second moments: TKE, TPE, and the vertical turbulent fluxes of the momentum and the buoyancy (proportional to the potential temperature). It takes into account the non-gradient correction to the down-gradient formulation for the vertical turbulent flux of buoyancy, and employs the concept of total turbulent energy (TTE = TKE + TPE). It is a model without a critical Richardson number permitting sustenance of turbulence by shear at any gradient Richardson number

$$\mathrm{Ri} = N^2/S^2. \tag{1}$$

Here, $N$ is the Brunt-Väisälä frequency defined as $N^2 = \beta \partial \Theta/\partial z$, $\Theta$ is the mean potential temperature, $\beta = g/T_0$ is the buoyancy parameter, $g$ is the acceleration of gravity, and $T_0$ is the reference value of the absolute temperature $T$. For the turbulent Prandtl number, defined as

$$\mathrm{Pr}_T = K_M/K_H, \tag{2}$$

where $K_M$ and $K_H$ are the eddy viscosity and eddy conductivity, the EFB model predicts the asymptotically linear dependence:

$$\mathrm{Pr}_T \propto \mathrm{Ri} \quad \text{at} \quad \mathrm{Ri} \gg 1. \tag{3}$$

In terms of the flux Richardson number, $\mathrm{Ri}_f$, and the Obukhov length scale, $L$, defined as

$$\mathrm{Ri}_f = \frac{-\beta F_z}{\tau S}, \tag{4}$$

$$L = \frac{\tau^{3/2}}{-\beta F_z}, \tag{5}$$

where $F_z$ is the vertical turbulent flux of potential temperature, and $\tau$ is the modulus of the vertical turbulent flux of momentum, Eq. (3) yields the following asymptotic formulae:

$$\mathrm{Ri}_f = \frac{\tau^{1/2}}{SL} \to \mathrm{Ri}_f^\infty \text{ at } \mathrm{Ri} \to \infty, \qquad (6)$$

where $\mathrm{Ri}_f^\infty$ is the maximal flux Richardson number. In the EFB closure, $\mathrm{Ri}_f^\infty$ is a universal constant ($\mathrm{Ri}_f^\infty < 1$) to be determined empirically. The model reveals a transitional interval, $0.1 < \mathrm{Ri} < 1$, separating the two turbulent regimes of essentially different nature: strong turbulence at $\mathrm{Ri} \ll 1$ and weak turbulence which transports momentum but is much less efficient in transporting heat at $\mathrm{Ri} > 1$.

Alternative new closure models with no Ri-critical also employ the TTE-budget equation but avoid the direct use of the budget equations for turbulent fluxes suggested by Mauritsen et al. (2007), and modification of their prior second-order turbulence closure by Canuto et al. (2008). L'vov et al. (2008) and L'vov and Rudenko (2008) have performed detailed analyses of the budget equations for the Reynolds stresses in the turbulent boundary layer (relevant to the strong turbulence regime) taking into consideration the dissipative effect of the horizontal heat flux explicitly, in contrast to Zilitinkevich et al. (2007) "effective-dissipation approximation". All three budget equations for TKE, TPE and TTE were considered earlier by Canuto and Minotti (1993), Elperin et al. (2002) and Cheng et al. (2002). The third-order vertical transports of TKE and TPE caused by IGW in the two-layer system, comprising the turbulence-dominated atmospheric boundary layer and the IGW-dominated free atmosphere, was included in a simple turbulence closure model by Zilitinkevich (2002).

## 2. Large-scale IGW in the stably stratified sheared flows

In the present paper we focus on the effect of large-scale IGW on the stably stratified turbulence and do not discuss small-scale IGW. Accordingly, we consider the IGW wavelength / periods much larger than the turbulence spatial / time scales. This allows us to treat the large-scale IGW with respect to turbulence as a kind of mean flows with random phases, and to neglect molecular dissipation of IGW. We also neglect the feedback effect of turbulence on IGW. At the low frequency part of the IGW spectra, we limit our analysis to frequencies essentially exceeding the Coriolis frequency, so that the IGW under consideration are not affected by the Coriolis parameter, $f = 2\Omega \sin\varphi$, where $\Omega_i$ is the Earth's rotation vector parallel to the polar axis ($|\Omega_i| \equiv \Omega = 0.76 \cdot 10^{-4}$ s$^{-1}$), and $\varphi$ is the latitude.

The large-scale IGW are characterized by the wave-field velocity, $\mathbf{V}^W = \left(V_1^W, V_2^W, V_3^W\right)$, and potential temperature, $\Theta^W$, which satisfy the following equations (in the Boussinesq approximation for incompressible fluid):

$$\frac{\partial \mathbf{V}^W}{\partial t} = -(\mathbf{U} \cdot \nabla) \mathbf{V}^W - \nabla\left(\frac{P^W}{\rho_0}\right) + \beta \Theta^W \mathbf{e} - (\mathbf{V}^W \cdot \nabla) \mathbf{V}^W, \tag{7}$$

$$\frac{\partial \Theta^W}{\partial t} = -(\mathbf{U} \cdot \nabla) \Theta^W - \frac{1}{\beta}(\mathbf{V}^W \cdot \mathbf{e}) N^2 - (\mathbf{V}^W \cdot \nabla) \Theta^W, \tag{8}$$

and the conditions of incompressibility: $\mathrm{div}\, \mathbf{V}^W = 0$ and $\mathrm{div}\, \mathbf{U} = 0$. Here, $\mathbf{U}$ is the mean flow velocity, $\beta = g/T_0$ is the buoyancy parameter, $g = 9.81$ m s$^{-1}$ is the acceleration due to gravity, $P^W$ is the pressure caused by IGW, $\mathbf{e}$ is the vertical unit vector, $\rho_0$ is the density of fluid, $N$ is the Brunt-Väisälä frequency: $N^2 = \beta \partial \Theta / \partial z$, $\Theta$ is the potential temperature defined as $\Theta = T(P_0/P)^{1-1/\gamma}$, $T$ is the absolute temperature, $T_0$ is its reference value, $P$ is the pressure, $P_0$ is its reference value, and $\gamma = c_p/c_v = 1.41$ is the specific heats ratio.

We do not consider nonlinear wave-wave interactions. Consequently, we neglect in Eqs. (7) and (8) the nonlinear terms $(\mathbf{V}^W \cdot \nabla) \mathbf{V}^W$ and $(\mathbf{V}^W \cdot \nabla) \Theta^W$, and apply to Eq. (7) the 'curl' operator to exclude the pressure ($P^W$) term. The solution of the linearised equations (7) and (8) in the Fourier space reads:

$$V_\alpha^W = -\frac{k_\alpha k_z}{k_h^2} V_0^W(\mathbf{k}) \cos(\omega t - \mathbf{k} \cdot \mathbf{r}), \tag{9}$$

$$\alpha = 1,2,$$

$$V_3^W \equiv V_z^W = V_0^W(\mathbf{k}) \cos(\omega t - \mathbf{k} \cdot \mathbf{r}), \tag{10}$$

$$\Theta^W = -\frac{Nk}{\beta k_h} V_0^W(\mathbf{k}) \sin(\omega t - \mathbf{k} \cdot \mathbf{r}) \tag{11}$$

(see, e.g., Turner 1973, Miropolsky 1981, Nappo 2002). Here, $\mathbf{k}$ is the wave vector; $k_\alpha = (k_x, k_y)$ is the horizontal wave vector, so that $k_h = \pm\sqrt{k_x^2 + k_y^2}$; and $\omega$ is the frequency of IGW:

$$\omega = \frac{k_h}{k} N + \mathbf{k} \cdot \mathbf{U}, \tag{12}$$

where $k = \sqrt{k_z^2 + k_h^2}$ is the total wave number. The second term in Eq. (12) is caused by the Doppler shift due to the sheared mean wind velocity $\mathbf{U}(z)$. Equations (9)-(10) satisfy the condition of incompressibility of the wave velocity field.

Propagation of IGW in the stably stratified sheared flows in the approximation of geometrical optics is determined by the following equations in the Hamiltonian form:

$$\frac{\partial \mathbf{r}}{\partial t} = \frac{\partial \omega}{\partial \mathbf{k}}, \qquad (13)$$

$$\frac{\partial \mathbf{k}}{\partial t} = -\frac{\partial \omega}{\partial \mathbf{r}}, \qquad (14)$$

where $\mathbf{r}$ is the radius-vector of the centre of the wave packet, and $\mathbf{k}$ is the characteristic wave-vector (see, e.g., Weinberg 1962). Since the Brunt-Väisälä frequency $N(z)$ and the mean velocity $\mathbf{U}(z)$ are the functions only of the vertical coordinate, $z$, i.e., the only non-zero spatial derivative in Eq. (14) is $\partial \omega / \partial z$, Eqs. (12) and (14) yield $\mathbf{k}_h = $ constant. For the Hamiltonian system of Eqs. (13)-(14), $d\omega/dt = 0$, and Eq. (12) yields

$$\frac{k_h}{k(z)} N(z) + \mathbf{k} \cdot \left( \mathbf{U}(z) - \mathbf{U}(Z_0) \right) = \frac{k_h}{k_0} N(Z_0), \qquad (15)$$

where $Z_0$ is the height of the IGW source, $k_0 = k(z = Z_0)$, and $\mathbf{k}_h(z) = \mathbf{k}_h(Z_0)$. We assume that the only source of the IGW is localised at $z = Z_0$ and neglect generation and dissipation of waves during their propagation in the atmosphere.

Equation (15) determining the $z$-dependence of $k(z) = \sqrt{k_z^2(z) + k_h^2}$ implies that the IGW vertical wave numbers, $k_z(z)$, change when the IGW propagates through the stably stratified sheared flow. For $Z_0 = 0$ (the IGW source is located at the surface), $k_z(z) = \sqrt{k^2(z) - k_h^2}$, and for $Z_0 = H$ (the IGW source is located at the upper boundary of the layer under consideration), $k_z(z) = -\sqrt{k^2(z) - k_h^2}$.

The IGW kinetic energy,

$$E_W \equiv \frac{1}{2} \int \left\langle \mathbf{V}_W^2(\mathbf{k}) \right\rangle_W d\mathbf{k} = \frac{1}{4} \int [V_0^W(\mathbf{k})]^2 \, d\mathbf{k}, \qquad (16)$$

is related to the energy spectrum $e_W(\mathbf{k})$ of the ensemble of IGW: $E_W = \int [e_W(\mathbf{k})/2\pi k^2] \, d\mathbf{k}$, and $\langle ... \rangle_W$ denotes time average over a IGW period. Here, the integration in the $\mathbf{k}$ space over the angle $\theta$ between the axis $z$ and the vector $\mathbf{k}$ is performed:
- from 0 to $\pi/2$ when the IGW source is located at the surface, and
- from $-\pi/2$ to 0, when the IGW source is located at the upper boundary of the layer.

Equations (9), (10) and (16) yield the expression for the wave amplitude: $[V_0^W(\mathbf{k})]^2 = (2k_h^2/\pi k^4) e_W(\mathbf{k})$. We assume that the energy spectrum of the ensemble of IGW generated at the point $Z_0$ is isotropic and has the power-law form:

$$e_W(k_0) = (\mu - 1) \, E_W \, H^{-(\mu+1)} \, k_0^{-\mu}, \tag{17}$$

where $E_W = \int [e_W(\mathbf{k}_0)/2\pi k_0^2] \, d\mathbf{k}_0 = \int e_W(k_0) \, dk_0$. Observations give different values of the exponent $\mu$ from 1 to 4 (Fofonoff, 1969; Pochapsky, 1972; Garrett and Munk, 1979; Miropolsky, 1981; Nappo 2002; Fritts and Alexander 2003). The wave vector $k_0$ varies from $H^{-1}$ to $L_W^{-1}$, where $L_W$ is the minimal wave length of the large-scale IGW. It is assumed that $L_W$ is much larger than the turbulence length scale but much smaller than the depth of fluid, $H$.

For simplicity we consider the power-law form, Eq. (17), of the energy spectrum of the ensemble of IGW. This standard assumption is supported by many experiments (e.g., Nappo 2002; Fritts and Alexander 2003; and references therein). Other forms of the energy spectrum would cause only minor changes in coefficients in the theoretical dependencies obtained below (in Sections 4-6) but would not change their form. The exponent $\mu$ is a free parameter, which must exceed unity and be less than 4 (see Nappo 2002; Fritts and Alexander 2003; and references therein). Variations in $\mu$ change only the coefficient on the r.h.s. of Eq. (43) for the IGW transport of momentum (see Section 4), and only weakly affect other theoretical dependencies.

## 3. Basic equations for turbulent flows accounting for large-scale IGW

We consider the large-scale IGW whose periods and wave lengths are much larger than the turbulent time and length scales. Therefore, although the IGW have random phases, the wave field interacts with the small-scale turbulence in the same way as the mean flow. We represent the total velocity as the sum of the mean-flow velocity, $\mathbf{U}(z)$, the wave-field velocity, $\mathbf{V}^W$, and the turbulent velocity, $\mathbf{u}$, and neglect the wave-wave interactions at large scales but take into account the turbulence-wave interactions. We limit our analysis to the flows, in which the vertical variations [along $x_3$ (or $z$) axis] of the mean wind velocity $\mathbf{U} = (U_1, U_2, U_3)$ and potential temperature $\Theta$ are much larger than the horizontal variations [along $x_1, x_2$ (or $x$, $y$) axes], so that the terms associated with the horizontal gradients in the budget equations for turbulent statistics can be neglected.

For typical atmospheric flows, the vertical scales (limited to the height scale of the atmosphere or the ocean: $H \sim 10^4$ m) are much smaller than the horizontal scales, so that the mean vertical velocity is much smaller than the horizontal velocity. To close the Reynolds equations in these conditions, we need only the vertical component, $F_z$, of the potential temperature flux and the two components of the vertical turbulent flux of momentum which comprise the turbulent contributions, $\tau_{13}$ and $\tau_{23}$, and the direct contributions of the large-scale IGW, $\tau_{1j}^{WW}$ and $\tau_{2j}^{WW}$.

The mean-flow momentum equations and thermodynamic energy equation accounting for the large-scale IGW can be written as follows:

$$\frac{DU_1}{Dt} = fU_2 - \frac{1}{\rho_0}\frac{\partial P}{\partial x} - \frac{\partial \tau_{13}}{\partial z} - \frac{\partial \tau_{1j}^{WW}}{\partial x_j}, \quad (18)$$

$$\frac{DU_2}{Dt} = -fU_1 - \frac{1}{\rho_0}\frac{\partial P}{\partial y} - \frac{\partial \tau_{23}}{\partial z} - \frac{\partial \tau_{2j}^{WW}}{\partial x_j}, \quad (19)$$

$$\frac{D\Theta}{Dt} = -\frac{\partial F_z}{\partial z} - \frac{\partial F_j^{WW}}{\partial x_j} + J, \quad (20)$$

where $D/Dt = \partial/\partial t + U_k \partial/\partial x_k$; $\tau_{ij} = \langle u_i u_j \rangle$; $F_i = \langle u_i \theta \rangle$; $t$ is the time; $\rho_0$ is the mean density; $J$ is the heating/cooling rate ($J = 0$ in adiabatic processes); $P$ is the mean pressure; $\mathbf{u} = (u_1, u_2, u_3) = (u, v, w)$ and $\theta$ are the velocity and potential-temperature fluctuations, respectively. The angle brackets $\langle ... \rangle$ denote the ensemble average over turbulent fluctuations. Besides the ensemble averaging, Eqs. (18)-(20) are averaged in time over the IGW period. This procedure is denoted by $\langle ... \rangle_W$. It implies that $\langle \cos(\omega t - \mathbf{k} \cdot \mathbf{r}) \rangle_W = \langle \sin(\omega t - \mathbf{k} \cdot \mathbf{r}) \rangle_W = 0$ and $\langle \cos^2(\omega t - \mathbf{k} \cdot \mathbf{r}) \rangle_W = 1/2$.

Direct effects of IGW on the mean flow are determined by the second-order moments

$$\tau_{\alpha j}^{WW} = \langle V_\alpha^W V_j^W \rangle_W, \quad (21a)$$
$$\alpha = 1, 2,$$
$$F_j^{WW} = \langle V_j^W \Theta^W \rangle_W, \quad (21b)$$

determining the wave-induced fluxes of momentum and potential temperature. In the linear theory, IGW do not transfer heat (so that $F_j^{WW} = 0$) but transfer momentum (e.g., Nappo, 2002). Accordingly, we neglect $F_j^{WW}$ but account for $\tau_{\alpha j}^{WW}$ (see Section 3.2).

The budget equations for the turbulent kinetic energy (TKE), $E_K = \frac{1}{2}\langle u_i u_i \rangle$, the squared potential temperature fluctuations, $E_\theta = \frac{1}{2}\langle \theta^2 \rangle$, and the potential-temperature flux, $F_i = \langle u_i \theta \rangle$, accounting for large-scale IGW can be written as follows:

$$\frac{DE_K}{Dt} + \frac{\partial \Phi_K}{\partial z} = -\tau_{i3} \frac{\partial U_i}{\partial z} + \beta F_z - \varepsilon_K - \left\langle \tau_{ij}^W \frac{\partial V_i^W}{\partial x_j} \right\rangle_W + \beta \left\langle V_z^W \Theta^W \right\rangle_W, \tag{22}$$

$$\frac{DE_\theta}{Dt} + \frac{\partial \Phi_\theta}{\partial z} = -F_z \frac{\partial \Theta}{\partial z} - \varepsilon_\theta - \left\langle F_j^W \frac{\partial \Theta^W}{\partial x_j} \right\rangle_W, \tag{23}$$

$$\frac{DF_i}{Dt} + \frac{\partial}{\partial x_j} \Phi_{ij}^{(F)} = \beta_i \langle \theta^2 \rangle + \frac{1}{\rho_0} \langle \theta \nabla_i p \rangle - \tau_{i3} \frac{\partial \Theta}{\partial z} - F_j \frac{\partial U_i}{\partial x_j} - \varepsilon_i^{(F)} - \left\langle \tau_{ij}^W \frac{\partial \Theta^W}{\partial x_j} \right\rangle_W - \left\langle F_j^W \frac{\partial V_i^W}{\partial x_j} \right\rangle_W. \tag{24}$$

Recall that $E_\theta$ is proportional to the turbulent potential energy (TPE):

$$E_P = \frac{\beta^2}{N^2} E_\theta, \tag{25}$$

so that Eq. (23) is equivalent to the budget equation for $E_P$.

As already mentioned, we are interested, first of all, in the vertical flux, $F_3 = F_z = \langle w\theta \rangle$, whose budget equation is

$$\frac{DF_z}{Dt} + \frac{\partial}{\partial z} \Phi_F = \beta \langle \theta^2 \rangle + \frac{1}{\rho_0} \left\langle \theta \frac{\partial}{\partial z} p \right\rangle - \langle w^2 \rangle \frac{\partial \Theta}{\partial z} - \varepsilon_z^{(F)} \\ - \left\langle \tau_{j3}^W \frac{\partial \Theta^W}{\partial x_j} \right\rangle_W - \left\langle F_j^W \frac{\partial V_z^W}{\partial x_j} \right\rangle_W. \tag{26}$$

Accounting for IGW, the budget equation for the Reynolds stresses, $\tau_{ij} = \langle u_i u_j \rangle$, reads:

$$\frac{D\tau_{ij}}{Dt} + \frac{\partial}{\partial x_k} \Phi_{ijk}^{(\tau)} = -\tau_{ik} \frac{\partial U_j}{\partial x_k} - \tau_{jk} \frac{\partial U_i}{\partial x_k} + \beta (F_j \delta_{i3} + F_i \delta_{j3}) + Q_{ij} - \varepsilon_{ij}^{(\tau)} \\ - \left\langle \tau_{ik}^W \frac{\partial V_j^W}{\partial x_k} \right\rangle_W - \left\langle \tau_{jk}^W \frac{\partial V_i^W}{\partial x_k} \right\rangle_W. \tag{27}$$

Hence, the budget equations for the non-diagonal, $\tau_{\alpha 3}$, and diagonal, $\tau_{\alpha\alpha} = 2E_\alpha$, components of the Reynolds stresses, $\tau_{ij} = \langle u_i u_j \rangle$, can be written as follows:

$$\frac{D\tau_{\alpha 3}}{Dt} + \frac{\partial}{\partial z}\Phi_{\alpha}^{(\tau)} = -\langle w^2 \rangle \frac{\partial U_{\alpha}}{\partial z} + \beta F_{\alpha} + Q_{\alpha 3} - \varepsilon_{\alpha 3}^{(\tau)} - \left\langle \tau_{\alpha j}^{W} \frac{\partial V_{z}^{W}}{\partial x_{j}} \right\rangle_{W} - \left\langle \tau_{j3}^{W} \frac{\partial V_{\alpha}^{W}}{\partial x_{j}} \right\rangle_{W}, \quad (28)$$

$$\frac{DE_{\alpha}}{Dt} + \frac{\partial}{\partial z}\Phi_{\alpha}^{(\tau)} = -\tau_{\alpha 3} \frac{\partial U_{\alpha}}{\partial z} + \frac{1}{2} Q_{\alpha\alpha} - \varepsilon_{\alpha\alpha}^{(\tau)} - \left\langle \tau_{\alpha j}^{W} \frac{\partial V_{\alpha}^{W}}{\partial x_{j}} \right\rangle_{W}, \quad (29)$$

$$\frac{DE_{z}}{Dt} + \frac{\partial}{\partial z}\Phi_{z}^{(\tau)} = \beta F_{z} + \frac{1}{2} Q_{33} - \varepsilon_{33}^{(\tau)} - \left\langle \tau_{j3}^{W} \frac{\partial V_{z}^{W}}{\partial x_{j}} \right\rangle_{W}, \quad (30)$$

where $\beta_i = \beta e_i$; $\mathbf{e} = (e_1, e_2, e_3)$ is the vertical unit vector; $\tau_{\alpha 3} = \langle u_{\alpha} w \rangle$ ($\alpha = 1, 2$) are the two components of the vertical turbulent flux of momentum, and $F_{\alpha} = \langle u_{\alpha} \theta \rangle$ are the horizontal fluxes of potential temperature ($\alpha = 1, 2$). In Eq. (29) we do not apply the summation convention for the double Greek indices.

The terms $\Phi_K$, $\Phi_{\theta}$ in Eqs. (22)-(23) are the third-order moments determining turbulent fluxes of $E_K$ and $E_{\theta}$:

$$\mathbf{\Phi}_K = \frac{1}{\rho_0} \langle p\, \mathbf{u} \rangle + \frac{1}{2} \langle u^2\, \mathbf{u} \rangle + \mathbf{\Phi}_K^W, \quad (31a)$$

and its $z$-component is

$$\Phi_K = \frac{1}{\rho_0} \langle p\, w \rangle + \frac{1}{2} \langle \mathbf{u}^2\, w \rangle + \Phi_K^W, \quad (31b)$$

$$\mathbf{\Phi}_{\theta} = \frac{1}{2} \langle \theta^2\, \mathbf{u} \rangle + \mathbf{\Phi}_{\theta}^W, \quad (31c)$$

and its $z$-component is

$$\Phi_{\theta} = \frac{1}{2} \langle \theta^2\, w \rangle + \Phi_{\theta}^W; \quad (31d)$$

where the terms marked with the superscript "$W$" denote the wave-driven turbulent fluxes of $E_K$ and $E_{\theta}$.

The terms $\Phi_{ij}^{(F)}$, $\Phi_F = \Phi_{33}^{(F)}, \Phi_{ijk}^{(\tau)}$ and $\Phi_{\alpha}^{(\tau)}$ in Eqs. (24)-(30) are the third-order moments representing the fluxes of fluxes:

$$\Phi_{ij}^{(F)} = \frac{1}{2\rho_0} \langle p\, \theta \rangle \delta_{ij} + \langle u_i\, u_j\, \theta \rangle + \Phi_{ij}^{(FW)}, \quad (32a)$$

$$\Phi_{33}^{(F)} = \Phi_F = \frac{1}{2\rho_0} \langle p\,\theta \rangle + \langle w^2\,\theta \rangle + \Phi_{33}^{(FW)}, \tag{32b}$$

$$\Phi_{ijk}^{(\tau)} = \langle u_i u_j u_k \rangle + \frac{1}{\rho_0} \left( \langle p u_i \rangle \delta_{jk} + \langle p u_j \rangle \delta_{ik} \right) + \Phi_{ijk}^{(\tau W)}, \tag{33a}$$

$$\Phi_i^{(\tau)} = \Phi_{i33}^{(\tau)} = \langle u_i w^2 \rangle + \frac{1}{\rho_0} \langle p u_i \rangle + \Phi_i^{(\tau W)}; \tag{33b}$$

where the terms marked with the superscript "(W)" denote the wave-driven turbulent fluxes of fluxes; and $Q_{ij}$ are correlations between the fluctuations of the pressure, $p$, and the velocity shears:

$$Q_{ij} = \frac{1}{\rho_0} \left\langle p \left( \frac{\partial u_i}{\partial x_j} + \frac{\partial u_j}{\partial x_i} \right) \right\rangle. \tag{34}$$

The terms $\varepsilon_K$, $\varepsilon_{ij}^{(\tau)}$, $\varepsilon_\theta$ and $\varepsilon_i^{(F)}$ are determined by the following formulas:

$$\varepsilon_K = \nu \left\langle \frac{\partial u_i}{\partial x_k} \frac{\partial u_i}{\partial x_k} \right\rangle, \tag{35a}$$

$$\varepsilon_{ij}^{(\tau)} = 2\nu \left\langle \frac{\partial u_i}{\partial x_k} \frac{\partial u_j}{\partial x_k} \right\rangle, \tag{35b}$$

$$\varepsilon_\theta = -\kappa \langle \theta \Delta \theta \rangle, \tag{36a}$$

$$\varepsilon_i^{(F)} = -\kappa \left( \langle u_i \Delta \theta \rangle + \Pr \langle \theta \Delta u_i \rangle \right), \tag{36b}$$

where $\nu$ is the kinematic viscosity, $\kappa$ is the temperature diffusivity, and $\Pr = \nu/\kappa$ is the Prandtl number.

The diagonal terms, $\varepsilon_{11}^{(\tau)}, \varepsilon_{22}^{(\tau)}, \varepsilon_{33}^{(\tau)}$, $\varepsilon_K$ (the sum of $\varepsilon_{ii}^{(\tau)}$), $\varepsilon_\theta$, and $\varepsilon_i^{(F)}$, representing the dissipation rates for $\tau_{\alpha\alpha}$, $E_K$, $E_\theta$ and $F_i^{(F)}$, respectively, are expressed using the Kolmogorov (1941) hypothesis:

$$\varepsilon_K = \frac{E_K}{C_K t_T}, \tag{37a}$$

$$\varepsilon_{\alpha\alpha}^{(\tau)} = \frac{\tau_{\alpha\alpha}}{C_K t_T}, \tag{37b}$$

$$\varepsilon_\theta = \frac{E_\theta}{C_P t_T}, \tag{37c}$$

$$\varepsilon_i^{(F)} = \frac{F_i}{C_F t_T}, \tag{37d}$$

where $t_T$ is the turbulent dissipation time scale; $C_K$, $C_P$ and $C_F$ are dimensionless constants; and the summation convention is not applied to the double Greek indices.

In the budget equations for the vertical turbulent fluxes of momentum, $\tau_{\alpha 3}$ ($\alpha = 1, 2$), the terms $\varepsilon_{\alpha 3}^{(\tau)}$ dependent on the molecular viscosity are usually small, whereas the contributions of the terms $\beta F_\alpha$ and $Q_{\alpha 3}$ to dissipation are overwhelming. Following Zilitinkevich et al. (2007), we introduce the Reynolds-stress "effective dissipation rates":

$$\varepsilon_{\alpha 3(\text{eff})} \equiv \varepsilon_{\alpha 3}^{(\tau)} - \beta F_\alpha - Q_{\alpha 3}, \tag{38}$$

$$\alpha = 1, 2;$$

and, by analogy with Eq. (37), apply to them the closure hypothesis:

$$\varepsilon_{\alpha 3(\text{eff})} = \frac{\tau_{\alpha 3}}{t_\tau} = \frac{\tau_{\alpha 3}}{C_\tau t_T}, \tag{39}$$

where $t_\tau$ is the effective dissipation time scale, and $C_\tau$ is a dimensionless coefficient accounting for the difference between $t_\tau$ and $t_T$. The turbulent dissipation time scale, $t_T$, is expressed through the vertical turbulent length scale, $l_z$, and the kinetic energy of the vertical velocity fluctuations:

$$t_T = \frac{l_z}{E_z^{1/2}}. \tag{40}$$

Equations (18)-(20) and (22)-(30) are obtained by averaging over the ensemble of turbulent fluctuations and over the period of large-scale IGW. These equations in a general form without the IGW terms can be found, e.g., in Kaimal and Finnigan (1994), Kurbatsky (2000), Cheng et al. (2002) and Canuto and Minotti (1993). Equation (22) is presented in Einaudi and Finnigan (1993). Hereafter we restrict our analysis to the effects of IGW on the second order statistics and leave the IGW third order moments (the fluxes of energies and the fluxes of momentum and heat fluxes) for further study.

The IGW terms in the above equations include the wave-field velocity and temperature, $V_i^W$ and $\Theta^W$, specified by Eqs. (9)-(11); and the instantaneous Reynolds stresses, $\tau_{ij}^W$, and turbulent flux of potential temperature, $F_i^W$, caused by the IGW-turbulence interaction. We determine $\tau_{ij}^W$ approximately – subtracting Eq. (27) from the ensemble-averaged equation for $\tau_{ij}$ but not averaged over the IGW period, assuming that $\omega t_T \ll 1$ and $\varepsilon_i^{(\tau)} = \tau_{ij}^W / (C_\tau t_T)$, and omitting the terms quadratic in wave amplitude, which do not contribute to the correlations $\langle \tau_{ij}^W (\partial V_i^W / \partial x_j) \rangle_W$ and $\langle \tau_{ij}^W (\partial \Theta^W / \partial x_j) \rangle_W$:

$$\tau_{ij}^W \approx -C_\tau t_T \left( \tau_{ik} \frac{\partial V_j^W}{\partial x_k} + \tau_{jk} \frac{\partial V_i^W}{\partial x_k} \right). \tag{41}$$

Similarly, we determine $F_i^W$ also approximately – subtracting Eq. (24) from the ensemble-averaged equation for $F_i$ but not averaged over the IGW period, assuming that $\omega t_T \ll 1$ and $\varepsilon_i^{(F)} = F_i^W / (C_F t_T)$, and omitting the terms quadratic in wave amplitude, which do not contribute to the correlations $\langle F_j^W (\partial V_i^W / \partial x_j) \rangle_W$ and $\langle F_j^W (\partial \Theta^W / \partial x_j) \rangle_W$:

$$F_i^W \approx -C_F t_T \left( \tau_{ij} \frac{\partial \Theta^W}{\partial x_j} + \tau_{i3}^W \frac{\partial \Theta}{\partial z} + F_j \frac{\partial V_i^W}{\partial x_j} \right). \tag{42}$$

Concrete effects of IGW on turbulence are considered in next sections.

## 4. The effects of large-scale IGW on the turbulent transports and energies

### 4.1. THE IGW TRANSPORT OF MOMENTUM

For simplicity, we consider the stationary, homogeneous regime of turbulence, neglect the effect of the Earth rotation, and assume that the mean wind velocity is directed along the $x$-axis: $\mathbf{U} = (U,0,0)$. Using Eqs. (9) and (10) for the IGW velocity field, Eq. (17) for the IGW energy spectrum, and assuming that $|U(z) - U(Z_0)| \ll L_W N(Z_0)$, integration over the spectrum of the IGW vertical flux of momentum, $\tau_{\alpha 3}^{WW}(\mathbf{k}_0) = -k_\alpha k_z(k_0) e_W(k_0) / [\pi k_0^2 k^2(k_0)]$, in $\mathbf{k}$-space yields:

$$\tau_{\alpha 3}^{WW} = \langle V_\alpha^W V_z^W \rangle_W = \int \tau_{\alpha 3}^{WW}(\mathbf{k}_0) d\mathbf{k}_0$$

$$= \pm \frac{\Gamma[U(Z_0) - U(z)]}{3N(z)H} [(Q+4)E(Q) - (Q+2)K(Q)] E_W, \tag{43}$$

where $Q$ is the dimensionless lapse rate:

$$Q = \left[\frac{N(z)}{N(Z_0)}\right]^2. \tag{44}$$

The coefficient $\Gamma$ is expressed through the exponent $\mu$ in the power-law energy spectrum of IGW, namely, for $1 < \mu < 2$:

$$\Gamma = \frac{\mu-1}{2-\mu}\left(\frac{H}{L_w}\right)^{2-\mu}, \tag{45a}$$

for $\mu = 2$:

$$\Gamma = \ln\left(\frac{H}{L_W}\right), \tag{45b}$$

for $2 < \mu$:

$$\Gamma = \frac{\mu-1}{\mu-2}, \tag{45c}$$

where $K(Q) = \int_0^{\pi/2}(1-Q^{-1}\sin^2\theta)^{-1/2}d\theta$ and $E(Q) = \int_0^{\pi/2}(1-Q^{-1}\sin^2\theta)^{1/2}d\theta$ are the complete elliptic integrals of the first and the second types, respectively. Plus or minus signs in Eq. (43) correspond to the cases when the IGW sources are located at the lower ($Z_0 = 0$) or upper ($Z_0 = H$) boundaries, respectively. The condition, $|U(z) - U(Z_0)| \ll L_W N(Z_0)$, is introduced to simplify further derivations and results, which otherwise become too cumbersome. This assumption is not principal and can be relaxed. At large $Q$, the integrals are $K(Q) = E(Q) \approx \pi/2$ and Eq. (43) reads:

$$\tau_{\alpha 3}^{WW} \approx \pm\frac{\pi}{3}\frac{\Gamma[U(z)-U(Z_0)]}{N(z)H}E_W. \tag{46}$$

When the IGW sources are located at the lower boundary ($Z_0 = 0$) and IGW are generated by the interaction of the flow with mountains or hills, $\tau_{\alpha 3}^{WW}$ is negative so that IGW transport momentum downward and increase the total downward momentum flux, $\tau_{\alpha 3} + \tau_{\alpha 3}^{WW}$ (where $\tau_{\alpha 3}^{WW} < 0$ and $\tau_{\alpha 3} < 0$). This known mechanism is called "wave drag" (e.g., Nappo, 2002). When the IGW sources are located at the upper boundary ($Z_0 = H$), e.g., when IGW propagating in the free atmosphere are trapped by the stably stratified atmospheric boundary layer (ABL), IGW transport momentum upwards ($\tau_{\alpha 3}^{WW} > 0$) because $U(z) < U(Z_0)$. Then the vertical flux of the momentum $\tau_{\alpha 3}^{WW}$ is subtracted from the turbulent flux, $\tau_{\alpha 3} < 0$, and the total vertical flux of momentum reduces. These effects can be parameterized using Eq. (43).

## 4.2. THE IGW PRODUCTION OF TURBULENT ENERGIES AND TURBULENT FLUX OF POTENTIAL TEMPERATURE

The IGW contribution to the production of TKE, $E_K$, is

$$\Pi^W = -\left\langle \tau_{ij}^W \frac{\partial V_i^W}{\partial x_j} \right\rangle_W = \int \Pi^W(\mathbf{k}_0) d\mathbf{k}_0, \qquad (47)$$

where $\Pi^W(\mathbf{k}_0)$ is the production of $E_K$ in **k**-space, caused by the diagonal components of the tensor $\tau_{jk}$:

$$\Pi^W(\mathbf{k}_0) = C_\tau t_T \tau_{jk} k_j k_k \frac{e_W(k_0)}{\pi k_0^2} = 2C_\tau t_T \left( E_z k_z^2(k_0) + E_x k_x^2 + E_y k_y^2 \right) \frac{e_W(k_0)}{\pi k_0^2}. \qquad (48)$$

Integration over **k** yields

$$\Pi^W = \frac{4C_\tau}{3} E_z^{1/2} l_z S^2 G \left[ \frac{1}{A_z} + 3(Q-1) \right], \qquad (49)$$

where $A_z = E_z/E_K$ is ratio of the vertical kinetic energy to TKE; $G$ is a dimensionless "wave-energy parameter" proportional to the normalized IGW kinetic energy, $E_W$:

$$G = \frac{E_W}{S^2 H^2} \frac{\mu - 1}{3 - \mu} \left( \frac{H}{L_W} \right)^{3-\mu}. \qquad (50)$$

In further analysis we assume that the wave lengths are much shorter than the basic depth scale: $L_w \ll H$.

The IGW contribution to the production of the vertical component of TKE, $E_z$, is

$$\Pi_z^w = -\left\langle \tau_{j3}^W \frac{\partial V_z^W}{\partial x_j} \right\rangle_W = \int \Pi_z^W(\mathbf{k}_0) d\mathbf{k}_0, \qquad (51)$$

where $\Pi_z^W(\mathbf{k}_0) = \Pi^W(\mathbf{k}_0) k_h^2 / k^2$ is the production of $E_z$ in **k**-space. Integration over **k** in Eq. (51) yields

$$\Pi_z^W = \frac{8C_\tau}{3} E_z^{1/2} l_z S^2 G \left[ 1 + \frac{2}{5Q}(A_z^{-1} - 3) \right]. \tag{52}$$

The IGW contribution to the productions of the longitudinal, $E_x$, and the transverse, $E_x$, components of TKE are

$$\Pi_x^W = \Pi_y^W = \frac{1}{2}\left( \Pi^W - \Pi_z^W \right) = \frac{2C_\tau}{3} E_z^{1/2} l_z S^2 G \left[ 3Q - 2 + (A_z^{-1} - 3)\left(1 - \frac{4}{5Q}\right) \right]. \tag{53}$$

The IGW contribution to the production of $E_\theta = \frac{1}{2}\langle \theta^2 \rangle$ is

$$\Pi_\theta^W = -\left\langle F_j^W \frac{\partial \Theta^W}{\partial x_j} \right\rangle_W = \int \Pi_\theta^W(\mathbf{k}_0) \, d\mathbf{k}_0, \tag{54}$$

where

$$\Pi_\theta^W(\mathbf{k}_0) = 2C_F t_T \frac{N^2(z)}{\beta^2} \left( E_z k_z^2(k_0) + E_x k_x^2 + E_y k_y^2 \right) \frac{e_W(k_0)}{\pi k_0^2}, \tag{55}$$

and $C_F$ is an empirical dimensionless constant. Here, we take into account that only diagonal components of the tensor $\tau_{jk}$ contribute to $\Pi_\theta^W(\mathbf{k}_0)$ (similarly to the production of $E_K$). Integration in Eq. (54) in **k**-space yields

$$\Pi_\theta^W = F_\theta^W \frac{N^2(z)}{\beta}, \tag{56}$$

where $F_\theta^W$ is the wave induced turbulent flux of potential temperature:

$$F_\theta^W = \frac{4C_F}{3\beta} E_z^{1/2} l_z S^2 G \left[ A_z^{-1} + 3(Q-1) \right]. \tag{57}$$

Note that the vertical flux of potential temperature is negative (downward), which is the reason for the negative production (see Eq. (56)). The IGW contribution to the production of $E_\theta$ is affected by $F_\theta^W$, which in its turn is affected by $E_\theta$.

In order to determine the direct effect of IGW on $F_\theta^W$, we take into account that $\langle V_i \Theta^W \rangle_W = 0$, which yields $\left\langle \tau_{ij}^W \dfrac{\partial \Theta^W}{\partial x_j} \right\rangle_W = 0$. Then, using Eq. (52), the term $\langle F_j^W (\partial V_z^W / \partial x_j) \rangle_W$ describing the production of $F_z$ in Eq. (26), can be written as follows:

$$\Pi_F^W = -\left\langle F_j^W \frac{\partial V_z^W}{\partial x_j} \right\rangle_W = -C_F \Pi_{33}^W \frac{l_z}{E_z^{1/2}} \frac{N^2(z)}{\beta}$$

$$= -\frac{8 C_F C_\tau}{3} l_z^2 S^2 G \left[ 1 + \frac{2}{5Q} (A_z^{-1} - 3) \right] \frac{N^2}{\beta} . \tag{58}$$

The production of the non-diagonal components of the Reynolds stresses, $\tau_{\alpha 3}$, caused by IGW is

$$\Pi_{\tau\alpha}^W = -\left\langle \tau_{\alpha j}^W \frac{\partial V_z^W}{\partial x_j} \right\rangle_W - \left\langle \tau_{j3}^W \frac{\partial V_\alpha^W}{\partial x_j} \right\rangle_W = \int \Pi_{\tau\alpha}^W (\mathbf{k}_0) d\mathbf{k}_0 , \tag{59}$$

where

$$\Pi_{\tau\alpha}^W (\mathbf{k}_0) = -2 C_\tau \, t_T \, \tau_{ij} \frac{k_i k_j k_\alpha k_z}{k^2(k_0)} \frac{e_W(k_0)}{\pi k_0^2} . \tag{60}$$

Here we take into account that only diagonal components of the tensor $\tau_{jk}$ contribute to $\Pi_{\tau\alpha}^W(\mathbf{k}_0)$. Using Eq. (17) and integrating in **k**-space in equation (59) yields:

$$\Pi_{\tau\alpha}^W = -\frac{8 C_\tau}{3} \tau_{\alpha 3} \frac{l_z S^2}{E_z^{1/2}} G \left( 1 - \frac{4}{5Q} \right) . \tag{61}$$

Finally, the production of the total turbulent energy (TTE) $E = E_K + E_P$ caused by IGW is

$$\Pi_E^W = \Pi^W + \Pi_\theta^W \frac{\beta^2}{N^2} . \tag{62}$$

Consequently, IGW contribute to the production of both TKE and TPE, in contrast to the mean shear, which contributes only to the TKE production.

# 5. Turbulence closures with and without IGW for the steady state regime

5.1. THE BACKGROUND ENERGY- AND FLUX-BUDGET (EFB) CLOSURE MODEL

In this section we present a refined version of the EFB turbulence closure model (Zilitinkevich et al., 2007). The latter employed the same equations as Eqs. (22)-(30) but without the IGW terms (marked in the present paper with the superscript "W"). Zilitinkevich et al. (2007) assumed that the dissipation constant for the kinetic and potential energies were equal ($C_P = C_K$) and had to admit that the ratio $C_\tau = t_\tau / t_T$ depends on Ri. Our analysis of the experimental data revealed that this assumption was not quite correct:
- the dissipation constants are different: $C_P/C_K = 0.72$,
- accounting for this difference, the coefficient $C_\tau$ turns into a universal (independent of Ri) constant.

This leads to essentially simplified EFB closure model – with $C_P \neq C_K$ but $C_\tau =$ constant, and yields a very simple formula for the eddy viscosity: $K_M = 2C_\tau E_z^{1/2} l_z$.

Note that principally the same result, $K_M (E_z^{1/2} l_z)^{-1} =$ constant, has been derived from quite rigorous analysis of the budget equations for the Reynolds stresses in **k**-space based on the $\tau$-approximation (Elperin et. al., 2002, 2006).

In order to better fit the EFB model to the available observational data on the vertical anisotropy, Zilitinkevich et al. (2007) proposed a modified formulation of the Rotta (1951) return-to-isotropy hypothesis. Considering the IGW-turbulence interaction, we now recognise that the apparent deviations from the Rotta hypothesis are caused by the effect of IGW. Therefore, it is only natural to retain the universally recognised classical formulation whereby the sum of the terms $\sum Q_{ii} = \sum \rho_0^{-1} \langle p \partial u_i / \partial x_i \rangle$ in Eqs. (28)-(30) is zero because of the continuity equation ($\sum \partial u_i / \partial x_i = 0$), so that $Q_{\alpha\alpha}$ describe the energy transfer from the high energy to the lower energy components:

$$Q_{\alpha\alpha} = -\frac{2C_r}{3C_K t_T}(3E_\alpha - E_K), \qquad (63)$$

where $C_r$ is a dimensionless constant accounting for the difference between the relaxation (return to isotropy) and dissipation time scales.

The terms $\beta \langle \theta^2 \rangle$ and $\rho_0^{-1} \langle \theta \partial p / \partial z \rangle$ in the budget equation (26) for the vertical turbulent flux of potential temperature play very important role. Zilitinkevich et al. (2007) showed that $\rho_0^{-1} \langle \theta \partial p / \partial z \rangle$ is negative and scales as $\beta \langle \theta^2 \rangle$, which yields the formula:

$$\beta\langle\theta^2\rangle + \rho_0^{-1}\langle\theta\,\partial p/\partial z\rangle = C_\theta\,\beta\langle\theta^2\rangle, \qquad (64)$$

where $C_\theta < 1$ is an empirical constant. We retain this approximation in the present study.

On these grounds we afford different values of $C_P$ and $C_K$ and essentially simplify the original EFB model setting $C_\tau$ = const and using the standard return-to isotropy formulation, Eq. (63).

## 5.2. THE EFB+IGW CLOSURE MODEL

Now we generalise the refined EFB closure model (Section 5.1) considering the budget equations (22)-(30) with the IGW terms determined by Eqs. (49), (52), (56)-(58). To demonstrate the role of IGW, we compare the two version of the closure – with and without IGW – *in the steady-state regime of turbulence*, when the left hand sides (l.h.s.) of all budget equations are zero, so that the model reduces to a system of algebraic equations. We focus on the turbulent energies and fluxes and leave the problem of determining the vertical turbulent length scale, $l_z$, for a separate study. In further derivations we basically follow Zilitinkevich et al. (2007) but introduce the changes indicated in Section 5.1 and include the effects of IGW presented in Sections 3 and 4.

In the steady state, the system of equations (22)- (23), (26), (28) and (30) reads:

$$-\tau_{i3}\frac{\partial U_i}{\partial z} + \beta F_z - \frac{E_K}{C_K\,t_T} + \Pi^W = 0, \qquad (65)$$

$$-F_z\frac{N^2}{\beta} - \frac{E_\theta}{C_P\,t_T} + \Pi_\theta^W = 0, \qquad (66)$$

$$2C_\theta\,\beta E_\theta - 2E_z\frac{N^2}{\beta} - \frac{F_z}{C_F\,t_T} + \Pi_F^W = 0, \qquad (67)$$

$$-2E_z\frac{\partial U_\alpha}{\partial z} - \frac{\tau_{\alpha 3}}{C_\tau\,t_T} + \Pi_{\tau\alpha}^W = 0, \qquad (68)$$

$$\beta F_z - \frac{C_r}{3C_K t_T}(3E_z - E_K) - \frac{E_z}{C_K\,t_T} + \Pi_z^W = 0, \qquad (69)$$

where the productions, $\Pi^W, \Pi_z^W\ \Pi_\theta^W, \Pi_F^W$ and $\Pi_{\tau\alpha}^W$, are determined by Eqs. (49), (52), (56), (58) and (61).

In the steady state, Equations (65)-(69) specify the turbulent energies and the vertical turbulent fluxes as dependent on the turbulent length scale, $l_z$, and the following dimensionless external parameters:
- gradient Richardson number Ri, Eq. (1),
- wave-energy parameter $G$, Eq. (50),
- lapse rate parameter $Q$, Eq. (44), characterising the IGW refraction.

A remarkable feature of this system is that $l_z$ drops out from the equations specifying the dimensionless parameters of turbulence[1], so that the latter are determined as universal functions of Ri, $G$ and $Q$, without any knowledge about $l_z$. In particular, for the dimensionless vertical TKE $\hat{E}_z$, the vertical flux of potential temperature $\hat{F}_z$ and the energy ratio $A_z$ defined as

$$\hat{E}_z \equiv \frac{E_z}{(S l_z)^2}, \tag{70a}$$

$$\hat{F}_z \equiv \frac{-\beta F_z}{E_z^{1/2} l_z S^2}, \tag{70b}$$

$$A_z \equiv \frac{E_z}{E_K}, \tag{70c}$$

the system reduces to the following three algebraic equations:

$$\hat{E}_z - \frac{2 C_K C_r C_\tau}{3(1+C_r)} \left\{ 1 - \left(1 + \frac{3}{C_r}\right) \frac{\hat{F}_z}{2 C_\tau} + G \left[ \frac{1}{3 A_z}\left(1 + \frac{12}{5 C_r Q}\right) + Q - 1 + \frac{2}{C_r}\left(1 - \frac{6}{5Q}\right) \right] \right\} = 0 \tag{71}$$

$$A_z - \frac{\hat{E}_z}{2 C_K C_\tau}\left(1 - \frac{\hat{F}_z}{2 C_\tau} + \frac{2G}{3}\left(A_z^{-1} + 3(Q-1)\right)\right)^{-1} = 0, \tag{72}$$

$$G\left[\frac{1}{A_z}\left(\frac{2 C_\tau}{5Q} - C\right) + C_\tau\left(1 - \frac{6}{5Q}\right) - 3C(Q-1)\right] + \frac{3\hat{E}_z}{4 C_F}\left[1 - \hat{F}_z\left(\frac{1}{2 C_F \text{Ri}} + \frac{C_\theta C_P}{\hat{E}_z}\right)\right] = 0, \tag{73}$$

where $C = C_\theta C_K$. The system of algebraic equations (71)-(73) determines the three functions:

$$\hat{E}_z = \hat{E}_z(\text{Ri}, G, Q), \tag{74a}$$
$$\hat{F}_z = \hat{F}_z(\text{Ri}, G, Q), \tag{74b}$$
$$A_z = A_z(\text{Ri}, G, Q), \tag{74c}$$

---

[1] This result is very favourable. It allows us to fully separate the problems of the energy and flux budgets (considered in this paper) and the vertical turbulent length scale, $l_z$ (to be considered later).

which can be found numerically. Other important dimensionless parameters of turbulence are expressed through $\hat{E}_z$, $\hat{F}_z$ and $A_z$:

$$\mathrm{Ri}_f \equiv \frac{-\beta F_z}{\tau S} = \frac{\hat{F}_z}{2C_\tau}\left[1 + \frac{8C_\tau^2}{3\hat{E}_z}G\left(1 - \frac{4}{5Q}\right)\right], \tag{75}$$

$$\left(\frac{\tau}{E_K}\right)^2 \equiv \left(\frac{\tau_{13}^2 + \tau_{23}^2}{E_K}\right) = \frac{\hat{E}_z}{C_K^2}\left(1 - \frac{\hat{F}_z}{2C_\tau} + \frac{2G}{3}\left(A_z^{-1} + 3(Q-1)\right)\right)^{-2}\left[1 + \frac{8C_\tau^2}{3\hat{E}_z}G\left(1 - \frac{4}{5Q}\right)\right]^{-2}, \tag{76}$$

$$\frac{F_z^2}{E_K E_\theta} = \frac{2C_\tau A_z}{C_P \mathrm{Pr}_T}\left[1 + \frac{4C_F}{3\hat{F}_z}G\left(A_z^{-1} + 3(Q-1)\right)\right]^{-1}, \tag{77}$$

$$\frac{E_P}{E_K} = \frac{C_P}{2C_\tau C_K}\left[\hat{F}_z + \frac{4C_F}{3}G\left(A_z^{-1} + 3(Q-1)\right)\right]\left(1 - \frac{\hat{F}_z}{2C_\tau} + \frac{2G}{3}\left(A_z^{-1} + 3(Q-1)\right)\right)^{-1}. \tag{78}$$

In contrast with Eqs. (76)-(78), the vertical turbulent fluxes of momentum and potential temperature essentially depend on $l_z$:

$$\tau_{\alpha 3} = -K_M \frac{\partial U_\alpha}{\partial z}, \tag{79a}$$

$$K_M = 2C_\tau E_z^{1/2} l_z \left[1 + \frac{8C_\tau^2}{3\hat{E}_z}G\left(1 - \frac{4}{5Q}\right)\right]^{-1}, \tag{79b}$$

$$F_z = -K_H \frac{\partial \Theta}{\partial z}, \tag{80a}$$

$$K_H = \frac{K_M \mathrm{Ri}_f}{\mathrm{Ri}}, \tag{80b}$$

where $\mathrm{Ri}_f$ is determined by Eq. (75).

At Ri << 1, the above dimensionless parameters have precisely the same asymptotic limits as in our new EFB closure without IGW:

$$\mathrm{Pr}_T \to \mathrm{Pr}_T^{(0)} = \frac{C_\tau}{C_F}, \tag{81}$$

$$A_z \to A_z^{(0)} = \frac{C_r}{3(1+C_r)}, \tag{82}$$

$$\left(\frac{\tau}{E_K}\right)^2 \to \frac{2C_\tau A_z^{(0)}}{C_K}, \tag{83}$$

$$\frac{F_z^2}{E_K E_\theta} \to \frac{2C_F A_z^{(0)}}{C_P}, \tag{84}$$

where the superscript "(0)" denotes Ri $\to$ 0.

## 6. Comparison of the EFB+IGW model with empirical data

Zilitinkevich et al. (2007) assumed that $C_P = C_K$ and determined the empirical coefficients $C_r$, $C_K$, $C_F$, $C_\tau$ (in *op. cit.* designated by $\Psi_\tau$), and $C_\theta$ by comparison of results from the model with data from field and laboratory experiments, large-eddy simulations (LES) and direct numerical simulations (DNS) related to the asymptotic regimes at Ri << 1 and Ri >> 1. For Ri << 1 they employed, in particular, the following estimates: $A_z^{(0)}$ = 1/6 [after laboratory experiments on the wall-bounded turbulence (L'vov et al., 2006) and DNS (Moser et al., 1999)], $(\tau/E_K)^{(0)}$ = 0.26 and $[F_z^2(E_K E_\theta)^{-2}]^{(0)}$ = 0.11 (after their own comparative analyses of different data). The superscript (0) in the above notations means "at Ri << 1".

As already mentioned, we basically follow Zilitinkevich et al. (2007) but no longer assume that $C_P = C_K$. Then using the well established empirical values of the turbulent Prandtl number, $\Pr_T^{(0)}$ = 0.8, and the von Karman constant, $k_u$ = 0.4, in the wall law $dU/dz = \tau^{1/2}(k_u z)^{-1}$ for the neutrally stratified surface layer (where $l_z \sim z$), the four constants are immediately obtained:

$$C_r = 3A_z^{(0)}(1-3A_z^{(0)})^{-1} = 1, \tag{85}$$

$$C_K = k_u (A_z^{(0)})^{1/2} \left[(\tau/E_K)^{(0)}\right]^{-3/2} = 1.2, \tag{86}$$

$$C_\tau = C_K (2A_z^{(0)})^{-1} (\tau^2/E_K^2)^{(0)} = 0.25, \tag{87}$$

$$C_F = C_\tau / \Pr_T^{(0)} = 0.31, \tag{88}$$

where we used Eqs. (81)-(84). Note that these estimates employ only data for neutrally stratified flows and therefore are equally relevant to the EFB and the EFB+IGW models because the IGW effects diminish at Ri << 1.

The maximal flux Richardson number for the flow without IGW, $\mathrm{Ri}_f^*$, and the ratio $C_P/C_K$ can be roughly estimated using Eq. (7) in Zilitinkevich et al. (2008) which is derived from the budget equations for the kinetic and potential turbulent energies:

$$\frac{E_P}{E} = \frac{(C_P/C_K)\mathrm{Ri}_f}{1+(C_P/C_K - 1)\mathrm{Ri}_f} . \tag{89}$$

Using the following values for the parameters $\mathrm{Ri}_f^\infty = 0.2$ and $(E_P/E)^\infty = 0.15$ (see the thick solid line in Fig. 5 representing a median for different kinds of empirical data), and Eq. (89) we obtained that

$$C_P/C_K = 0.72 . \tag{90}$$

Because of lack of better data, we consider this value of $C_P/C_K$, relevant to the regime without IGW.

It remains to determine the constant $C_\theta$ in Eq. (67). In the EFB closure without IGW, it is expressed through the limiting values of the energy ratio, $A_z$, and the flux Richardson number, $\mathrm{Ri}_f$, at $\mathrm{Ri} \to \infty$. Then, adopting reasonable values, $\mathrm{Ri}_f^\infty = 0.2$ and $A_z^\infty = 0.056$, for the regime without IGW (solid lines in Figures 2-5 based on the DNS, LES and lab experiments presumably unaffected by IGW), we obtain

$$C_\theta = \frac{A_z^\infty}{\mathrm{Ri}_f^\infty}(1-\mathrm{Ri}_f^\infty) = 0.31, \tag{91}$$

where the superscripts "(0)" and "$\infty$" denote "at Ri = 0" and "at Ri $\to \infty$", respectively.

Typical atmospheric values of the wave-energy parameter $G$, Eq. (50), and the lapse rate parameter $Q$, Eq. (44), determining the effects of IGW in our closure model, are estimated as follows. The first parameter, $G$, is obviously non-negative and in the Earth's troposphere could vary from zero (in the absence of waves) to about 10 – in the layers with strong wave activity. In the stratosphere $G$ could be much larger, and IGW could become the major source of turbulence.

Since the IGW are trapped by the strongly stratified layers, we consider $N(z) \geq N(Z_0)$, that is $Q \geq 1$. Furthermore, the static stability of the troposphere varies only slightly around typical value of the Brunt-Väisälä frequency $N \approx 10^{-2}$ s$^{-1}$. Therefore, the reasonable estimates for the orographically generated IGW are: $N(z) \approx N(Z_0) \approx 10^{-2}$ s$^{-1}$, and $Q = [N(z)/N(Z_0)]^2 \approx 1$. In the alternative case, when IGW propagating in the free troposphere are trapped by the stronger stratified long-lived stable PBL, where $N^2 \approx 5(\beta F_z/\tau)^2$ s$^{-2}$ (e.g., Zilitinkevich and Esau, 2007), $Q$ could be a few times larger. A reasonable meteorological range of the lapse rate parameter is $1 < Q < 5$.

In the EFB+IGW model the maximal flux Richardson number $\text{Ri}_f^{\max} \equiv (\lim \text{Ri}_f \text{ at } \text{Ri} \to \infty)$ is no longer a universal constant. Its variations are controlled by the counteraction of the direct and indirect mechanisms of generation of the turbulent flux of potential temperature by IGW, namely, by the two terms on the right hand side (r.h.s.) of Eq. (26): "direct": $\Pi_F^w = -\langle F_j^w (\partial V_z^w / \partial x_j) \rangle_w$ and "indirect", caused by the temperature fluctuations: $C_\theta \beta \langle \theta^2 \rangle = \beta \langle \theta^2 \rangle + \rho_0^{-1} \langle \theta (\partial p / \partial z) \rangle$ [where $\langle \theta^2 \rangle$ satisfies Eq. (23)]. As shown in Figure 1, $\text{Ri}_f^{\max}$ with increasing $G$ increases at $1 \leq Q < 1.02$, and decreases at $Q > 1.03$. Note that the maximal flux Richardson number $\text{Ri}_f^{\max}$ at $Q > 1.03$ reaches zero at some value of $G$ (dependent on $Q$). For larger $G$ the vertical flux of potential temperature becomes positive, that is counter-gradient. This looks surprising, but in fact is only natural. Indeed, IGW generate the potential temperature fluctuations, which in turn generate the upward (positive) contribution to the flux of potential temperature (cf. the above "indirect" mechanism). When the "indirect" share of the flux becomes larger than the "direct" share, the resulting flux changes the sign and becomes positive in spite of the stable stratification.

Figures 2-5 show empirical data on the Ri-dependences of the turbulent Prandtl number, $\text{Pr}_T$, flux Richardson number, $\text{Ri}_f$, energy-flux ratios, $(E_K / \tau)^2$ and $(E_K E_\theta) / F_z^2$, and energy ratios $A_z$ and $E_P / E$ together with theoretical curves plotted after the EFB model (heavy solid lines) and the EFB+IGW model for different $G$ and $Q$. Since at $\text{Ri} < 0.25$, large-scale IGW practically do not affect turbulence, the model predictions accounting for IGW are plotted in Figs. 2-5 only for $\text{Ri} > 0.25$.

Recall that we consider the simplest version of our closure model relevant to the stationary homogeneous regime of turbulence (with no non-local sources of turbulent energies or turbulent fluxes). On the contrary, most of available empirical data represent vertically (and in some cases also horizontally) heterogeneous flows, controlled (besides Ri, $G$ and $Q$) by additional, practically unavailable parameters. In this context, empirical Ri dependencies of $\text{Pr}_T$, $\text{Ri}_f$, $(\tau / E_K)^2$, $F_z^2 / (E_K E_\theta)$, $A_z$ and $E_P E_K^{-1}$ demonstrated by Mauritsen and Svensson (2007) and Zilitinkevich et al. (2007; 2008) are encouraging.

Below we attempt to more accurately determine empirical constants of the model. For this purpose, we rule out data suspicious for strong heterogeneity, and limit our analyses to meteorological data of Kondo et al. (1978), Bertin et al. (1997), Banta et al. (2002), Poulos et al. (2002), Uttal et al. (2002) and Mahrt and Vickers (2005); laboratory data of Strang and Fernando (2001), Rehmann and Koseff (2004) and Ohya (2001); LES data of (Esau 2004) and Zilitinkevich et al. (2008); and DNS data of Stretch et al. (2001).

Empirical Ri-dependencies of the turbulent Prandtl number, $\text{Pr}_T$, and flux Richardson number, $\text{Ri}_f$, are shown in Figure 2 together with the two kinds of theoretical curves: heavy solid lines calculated neglecting IGW for $\text{Ri}_f^\infty = 0.2$; and bunches of thin lines calculated accounting for IGW (for different $G$ and $Q$) in the interval $\text{Ri} \geq 0.25$. The latter cover the range of variability of presented data, which allows us to at least partially attribute the spread of data to the IGW mechanisms. The same format is used to show the energy to flux ratios: $(E_K/\tau)^2$ and $(E_K E_\theta)/F_z^2$ in Figure 3; the energy ratios: $A_z = E_z/E_K$ and $E_P/E$ in Figures 4 and 5.

With increasing $G$, the theory predicts that $(E_K/\tau)^2$ and $(E_K E_\theta)/F_z^2$ increase and $A_z$ decreases. This looks only natural: in contrast to the mean shear, generating only horizontal velocity fluctuations, IGW generate both horizontal and vertical fluctuations and directly contribute to $A_z$. Similarly, the energy ratio, $E_P/E$, increases with increasing $G$ due to direct generation of turbulent potential energy by IGW. It is worth mentioning that Figure 3b reveals the linear asymptote: $(E_K E_\theta)/F_z^2 \sim \text{Ri}$ at $\text{Ri} \gg 1$.

## 7. Concluding remarks

In the stably stratified atmospheric and oceanic flows, large-scale IGW directly perform vertical transport of momentum and contribute the TKE and TPE generation. Furthermore, the mean squared potential temperature fluctuation, $\langle \theta^2 \rangle$ proportional to the TPE, essentially controls generation of the vertical turbulent flux of potential temperature, which is why this flux is also affected by IGW.

In contrast to the mean shear, which directly generates only the horizontal component of TKE, large-scale IGW generate all three TKE components: $E_x$, $E_y$ and $E_z$, and therefore essentially reduce anisotropy, that is increase the parameter $A_z = E_z/E_K$. This effect is especially pronounced in very stable stratification and quite probably represents the key source of a very large scatter in empirical plots of $A_z$ versus Ri.

Furthermore, large-scale IGW generate both kinetic and potential turbulent energies and, as a rule, increase the share of the potential energy. Consequently, the maximal flux Richardson number ($\text{Ri}_f^{\max}$ attainable at $\text{Ri} \to \infty$) is no longer a universal constant ($\text{Ri}_f^\infty$) as it was in the EFB model, but a variable parameter essentially dependent on both the IGW energy parameter $G$, Eq. (50), and the lapse rate parameter $Q$, Eq. (44). At different $Q$, this effect causes larger as well as smaller values of $\text{Ri}_f^{\max}$ (see Figure 1).

At $Q < 1.03$, the theory leaves room for the values of $\text{Ri}_f^{\max}$ exceeding 1 – obviously impossible in the stationary homogeneous flows without IGW, but observed in some experiments. On the contrary, at $Q > 1.03$ and sufficiently large values of the wave energy parameter, $G$, the maximal flux Richardson number, $\text{Ri}_f^{\max}$, reaches zero and then becomes negative, which means that the vertical flux of potential temperature, $F_z$, becomes positive, in spite of $\beta \partial \Theta / \partial z \equiv N^2 > 0$. The point is that IGW directly produce potential temperature fluctuations, which in turn produce the upward contribution to $F_z$. When it exceeds the contribution due to the potential temperature gradient, the resulting flux changes the sign and becomes counter-gradient.

When the sources of IGW are located at the lower boundary of the air flow ($Z_0 = 0$), in particular, when IGW are generated by the flow interaction with mountains or hills, the vertical flux of momentum caused by IGW, $\tau_{\alpha 3}^{WW}$, is negative and contributes to the total (turbulent + wave induced) flux: $\tau_{\alpha 3} + \tau_{\alpha 3}^{WW} < 0$. This known mechanism is called "wave drag" (e.g., Nappo, 2002).

When the source of IGW is located at the upper boundary of a strongly stratified atmospheric boundary layer trapping the IGW from the free atmosphere, $Z_0$ can be identified with the boundary-layer height. Then the velocity difference $U(z) - U(Z_0)$ is negative; and the wave-induced vertical flux of momentum, $\tau_{\alpha 3}^{WW}$, determined by Eq. (43) is oriented upwards: $\tau_{\alpha 3}^{WW} > 0$. It follows that $\tau_{\alpha 3}^{WW}$ counteracts the ordinary vertical turbulent flux of momentum, $\tau_{\alpha 3} < 0$; so that the total momentum flux and therefore the level of turbulence in the ABL diminish.

To the best of our knowledge, the above mentioned IGW mechanism leading to the counter-gradient heat transfer at large positive gradient Richardson numbers, as well as the upward transfer of momentum and consequent weakening of the boundary-layer turbulence by trapped IGW, have not been considered until present. The trapped-wave effect could form the basis for dangerous air pollution events and is therefore of practical interest.

It goes without saying that the above unexpected theoretical predictions call for empirical verification. Empirical constants of our turbulence closure model, the most important of which is $\text{Ri}_f^\infty$, also need to be more carefully determined from field and laboratory experiments, DNS and LES.

## Acknowledgements

This work has been supported by the EC FP7 projects ERC PBL-PMES (No. 227915) and MEGAPOLI (No. 212520), and the Israel Science Foundation governed by the Israel Academy of Sciences (grant No. 259/07).

# Figure captions

**Figure 1.** Maximal values $\text{Ri}_f^{\max}$ (attainable at $\text{Ri} \to \infty$) of the flux Richardson number, $\text{Ri}_f = -\beta F_z (\tau S)^{-1}$, as functions of the wave energy parameter, $G$, Eq. (50), for different values of the lapse rate parameter, $Q$, Eq. (44).

**Figure 2.** Ri-dependences of (a) turbulent Prandtl number, $\text{Pr}_T = K_M / K_H$, and (b) flux Richardson number, $\text{Ri}_f$. Data points show *meteorological observations*: slanting black triangles (Kondo et al., 1978), snowflakes (Bertin et al., 1997); *laboratory experiments*: black circles (Strang and Fernando, 2001), slanting crosses (Rehmann and Koseff, 2004), diamonds (Ohya, 2001); *LES*: triangles (Zilitinkevich et al., 2008); and *DNS*: five-pointed stars (Stretch et al., 2001). Curves are plotted after our model (with $\text{Ri}_f^\infty = 0.2$): thick solid lines for the no-IGW regime ($G = 0$); thin dashed-dotted lines for $Q = 1$ and $G = 5$; dashed lines for $Q = 1$ and $G = 1$; thick dashed-dotted lines for $Q = 1$ and $G = 0.5$; thin solid lines for $Q = 1.5$ and $G = 0.3$; dotted lines for $Q = 1.5$ and $G = 0.2$.

**Figure 3.** Same as in Figure 2 but for the energy to flux ratios: (a) $(E_K / \tau)^2$ and (b) $(E_K E_\theta) / F_z^2$. Data points show *meteorological observations*: squares [CME = Carbon in the Mountains Experiment, Mahrt and Vickers (2005)], circles [SHEBA = Surface Heat Budget of the Arctic Ocean, Uttal et al. (2002)], overturned triangles [CASES-99 = Cooperative Atmosphere-Surface Exchange Study, Poulos et al. (2002), Banta et al. (2002)]; *laboratory experiments*: diamonds (Ohya, 2001); and *LES*: triangles (Zilitinkevich et al. (2008). Thick solid lines show the no-IGW version of the model ($G = 0$). Other lines are in Fig. 3a: thin dashed-dotted lines for $Q = 1$ and $G = 8$, dashed lines for $Q = 1$ and $G = 1$, thick dashed-dotted lines for $Q = 1$ and $G = 0.5$, thin solid lines for $Q = 1.5$ and $G = 0.3$, dotted lines for $Q = 1.5$ and $G = 0.2$; and in Fig. 3b: thick dashed lines for $Q = 1$ and $G = 0.4$, thick dashed-dotted lines for $Q = 1$ and $G = 0.1$, thin dashed lines for $Q = 1.5$ and $G = 0.1$, thin dashed-dotted lines for $Q = 1.5$ and $G = 0.05$.

**Figure 4.** Same as in Figures 2 and 3 but for the energy ratio $A_z = E_z / E_K$ with additional DNS data of Stretch et al. (2001) shown by five-pointed stars. The theoretical curve for the no-IGW regime ($G = 0$) is shown by the thick solid line; other theoretical curves are: thin dashed-dotted, for $Q = 1$ and $G = 5$; thick dashed, for $Q = 1$ and $G = 1$; thick dashed-dotted, for $Q = 1$ and $G = 0.5$; thin dotted, for $Q = 1$ and $G = 0.05$; thin dashed, for $Q = 1$ and $G = 0.01$; thin solid, for $Q = 1.5$ and $G = 0.3$; thick dotted, for $Q = 1.5$ and $G = 0.2$.

**Figure 5.** The ratio of the potential to total turbulent energies, $E_P / E$, versus the gradient Richardson number, Ri. Data points show *meteorological observations*: overturned triangles [CASES-99 = Cooperative Atmosphere-Surface Exchange Study, Poulos et al. (2002), Banta et al. (2002)]; *laboratory experiments*: diamonds (Ohya, 2001); and *LES*: triangles (Zilitinkevich et al. (2008). Curves are plotted after our model with $\text{Ri}_f^\infty = 0.2$: thick solid line for the no IGW

regime ($G = 0$); thin dashed-dotted line for $Q = 1$ and $G = 5$; thick dashed line for $Q = 1$ and $G = 1$; thick dashed-dotted line for $Q = 1$ and $G = 0.5$; thin dashed line for $Q = 1$ and $G = 0.1$; thin solid line for $Q = 1.5$ and $G = 0.3$; dotted line for $Q = 1.5$ and $G = 0.2$.

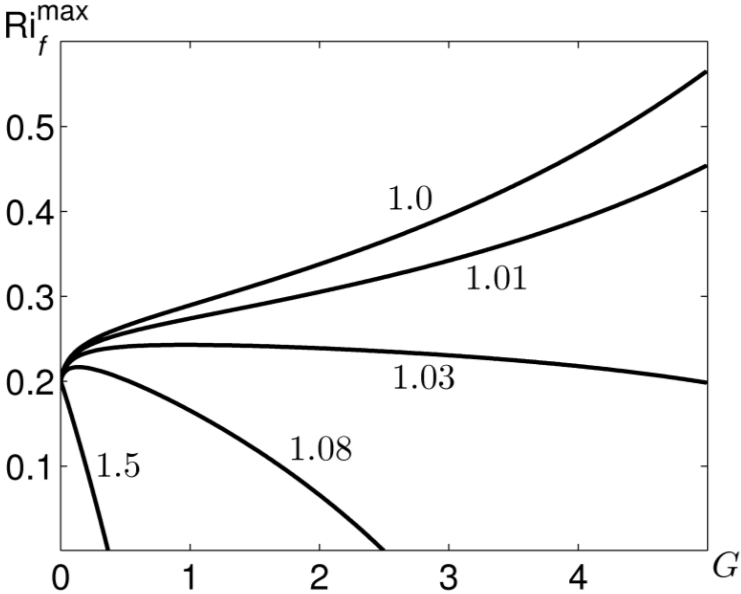

Figure 1.

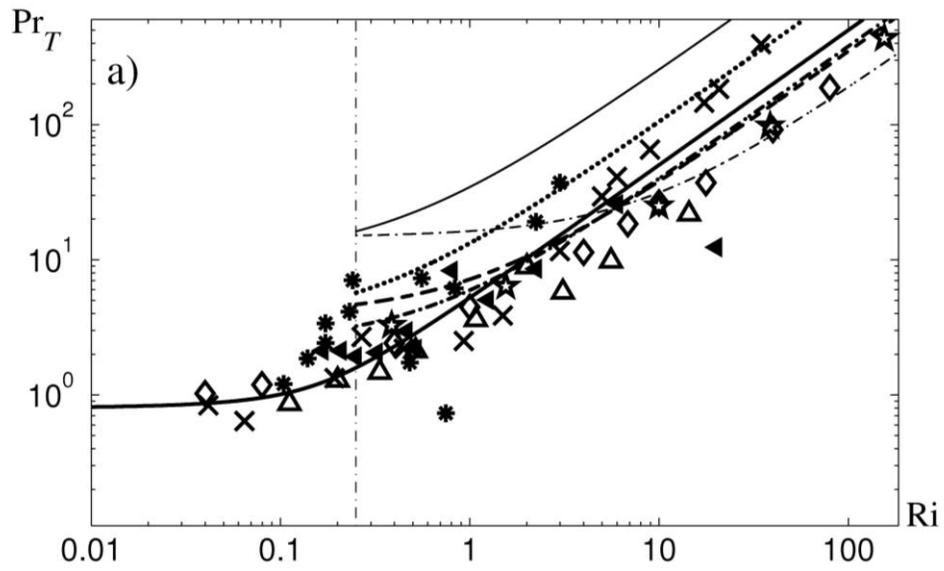
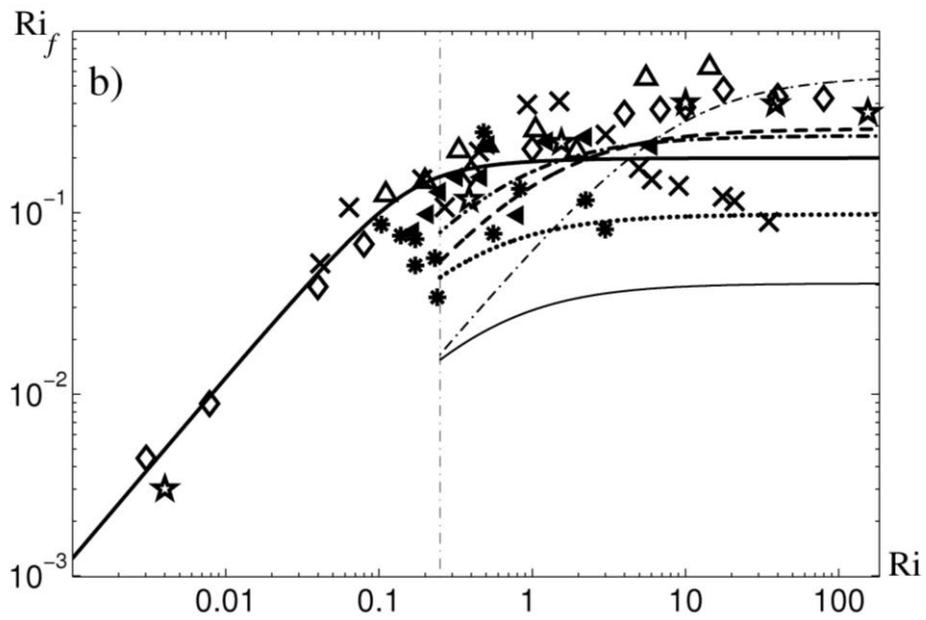

Figure 2.

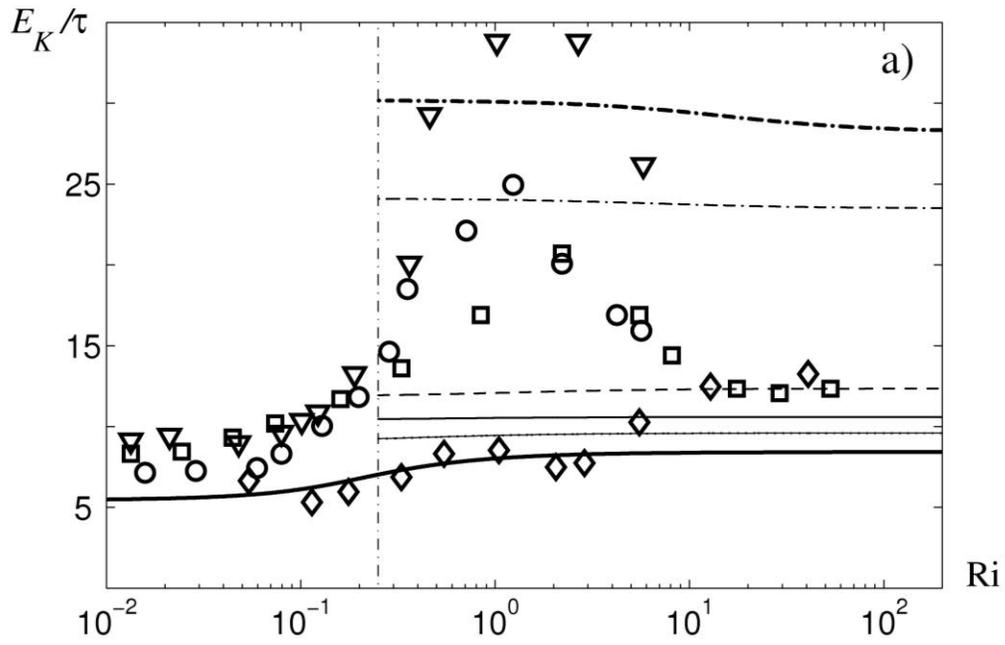
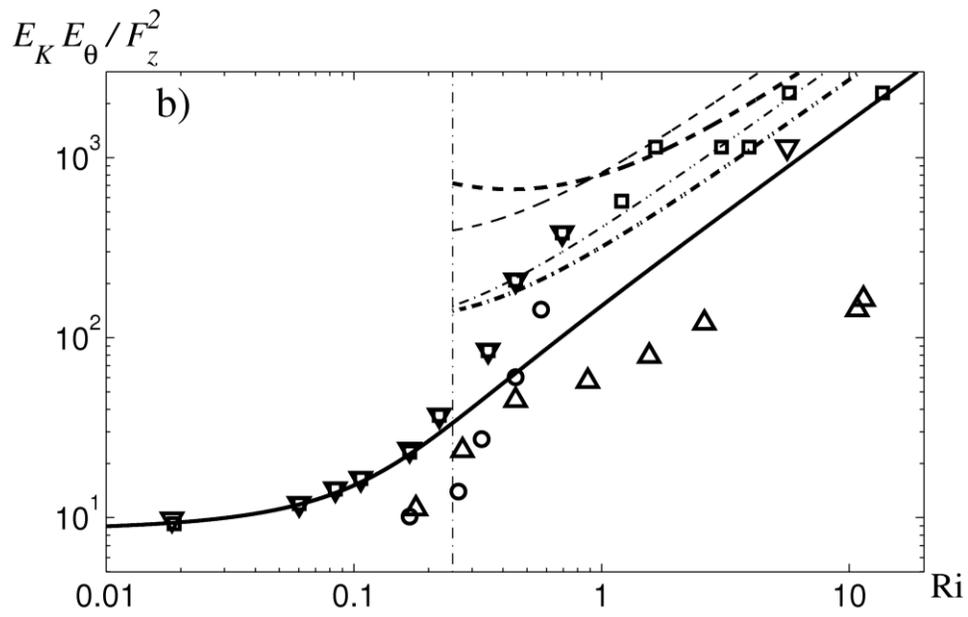

Figure 3.

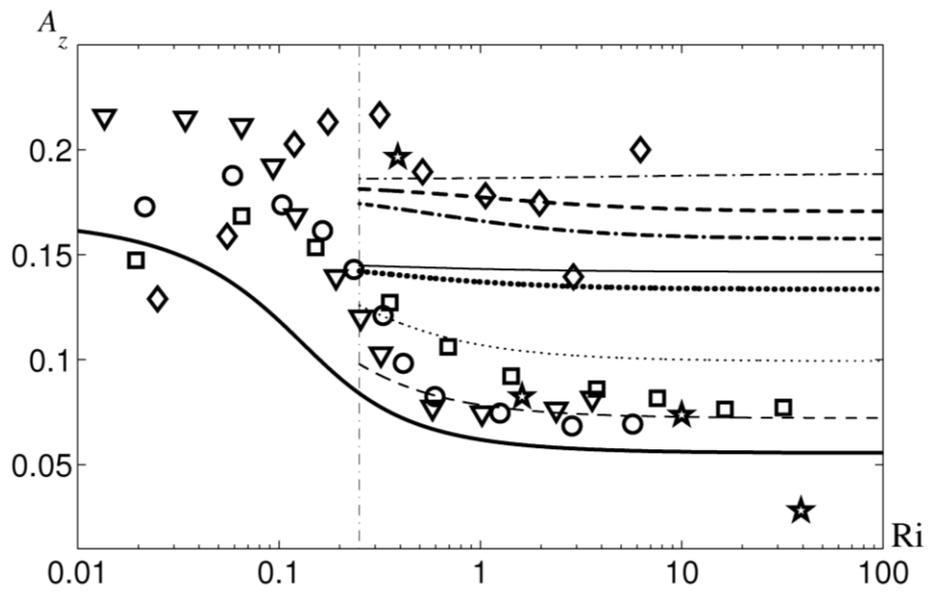

Figure 4

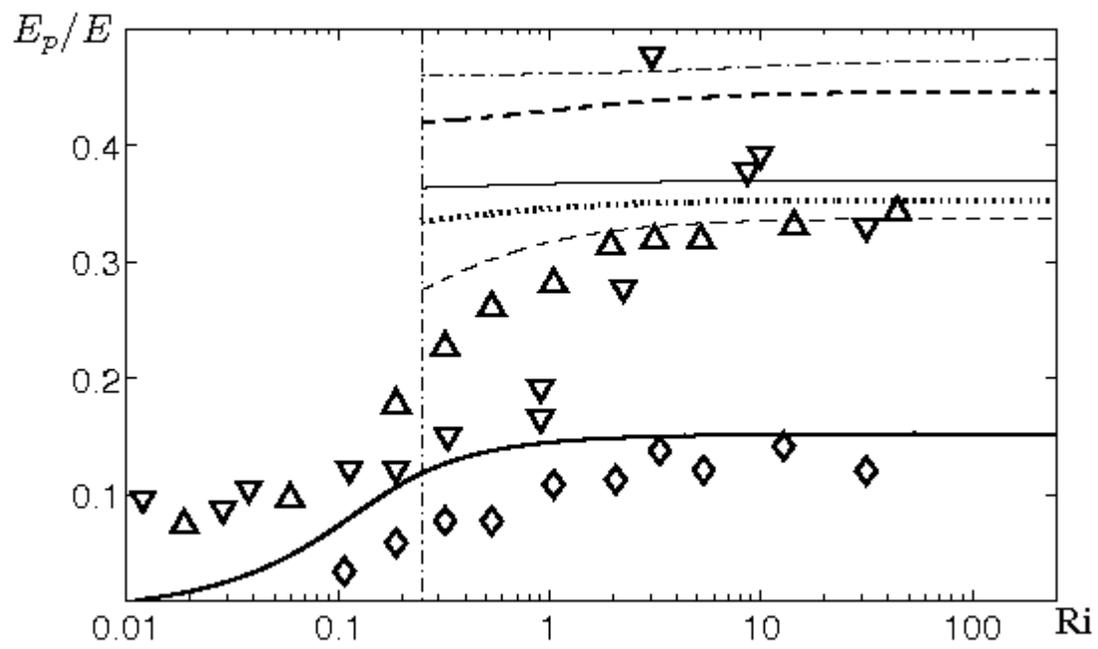

Figure 5